\documentclass[11pt]{article}
\usepackage{graphicx}
\begin{document}
%

\begin{center}
{\large \bf The Family Collider}

\vskip.5cm

W-Y. Pauchy Hwang\footnote{Correspondence Author;
 Email: wyhwang@phys.ntu.edu.tw}
 \\
{\em Asia Pacific Organization for Cosmology and Particle Astrophysics, \\
Institute of Astrophysics, Center for Theoretical Sciences,\\
and Department of Physics, National Taiwan University,
     Taipei 106, Taiwan}
\vskip.2cm


{\small(September 22, 2014)}
\end{center}

\begin{abstract}
Granting that the $SU_c(3) \times SU_L(2) \times U(1) \times SU_f(3)$
Standard Model is valid (or, partially valid), for the real world,
we propose the $\mu^+ e^-$ collider in the $10^2\, GeV$ range as the
family collider. This family collider may work efficiently in
producing the family Higgs particles and detecting the effects of
family gauge bosons, with the range of sub-sub-fermi's (a few $10^{-2}
\, fermi$'s).

\bigskip

{\parindent=0pt PACS Indices: 12.60.-i (Models beyond the standard
model); 98.80.Bp (Origin and formation of the Universe); 12.10.-g
(Unified field theories and models).}
\end{abstract}

\bigskip

\section{Introduction}

This is the experimental collider project dreamed by a theoretical
physicist.

Imagine that the lepton world is also governed by the family forces
and family particles, but the quark world does not feel it. Family
forces are assumed in the sub-sub-fermi range ($10^{-2}\, fermi$),
which is so much shorter than the mutual distances between two
electrons in the same atom. Family Higgs particles may be around
$(80-120\, GeV)$, and which could convert an electron into a muon.

Noting that $SU_c(3)$ couples alone to the quark world for no
reasons, $SU_f(3)$ is assumed to couples alone to the lepton
world - making the asymptotical freedom everywhere and escaping
the QED Landau's ghost for all particles.

The dreamed collider is based upon the $(\mu^+ e^-)$ collisions
at the center-of-mass energies of $(80-120)\, GeV$ - a collider
based on two well-controlled beams (so far).

\medskip

\section{The Basic Physics}

There were many speculations about the origin of mass; among them, the one in
the (old) Standard Model \cite{Book}, though "ugly", might be the closest
one in the thinking. Along this line, the author worked out the origin
of mass \cite{Origin} upon introducing the massive family gauge bosons.

In \cite{Origin}, it is demonstrated that, before spontaneous symmetry
breaking (SSB) in the generalized family Higgs mechanism, the system does
not have mass terms; upon SSB, every mass term appears as a result. This
works for masses of the various Higgs, of the leptons, of the quarks,
and of the various gauge bosons.

Early on, we propose\cite{HwangYan} $((\nu_\tau,\,\tau)_L,\,(\nu_\mu,\,\mu)_L,
\,(\nu_e,\,e)_L)$ $(columns)$ ($\equiv \Psi(3,2)$) as the
$SU_f(3)$ triplet and $SU_L(2)$ doublet. In fact, this is a natural proposal
so long as the idea of the family gauge theory can be adopted. Basically,
we introduce another $SU(3)$ that covers the lepton world and protects
it from the QED Landau ghost (and makes it asymptotically free). Note that
in notation we put $(\nu_\tau, \,\tau)$ as the first member of
$SU_f(3)$ to emphasize the family effects.

In writing out the $SU_c(3) \times SU_L(2) \times U(1) \times
SU_f(3)$ Standard Model in detail \cite{Hwang417}, the neutrino
mass term assumes a unique form:
\begin{equation}
i {h\over 2} {\bar\Psi}_L(3,2) \times \Psi_R(3,1) \cdot \Phi(3,2)
+ h.c.,
\end{equation}
where $\Psi(3,i)$ is the neutrino triplet just mentioned above (with
the first label for $SU_f(3)$ and the second for $SU_L(2)$). The
cross-dot (curl-dot) product is somewhat new, referring to the singlet
combination of three triplets in $SU(3)$. The Higgs field $\Phi(3,2)$ is
new in this effort \cite{HwangYan}, because it carries some nontrivial
$SU_L(2)$ charge.

Indeed, this off-diagonal neutrino mass term offers us a natural way to
describe neutrino oscillations, since the neutral part of $\Phi(3,2)$
could each receive vacuum expectation values.

On the other hand, for charged leptons, the Standard-Model choice is
$\Psi^\dagger(\bar 3,2) \Psi_R^C(3,1) \Phi(1,2) +c.c.$, which gives
three leptons an equal mass; but, in
view of that if $(\phi_1,\phi_2)$ is an $SU(2)$ doublet then $(\phi_2^\dagger,
-\phi_1^\dagger)$ is another doublet, we could form ${\tilde\Phi}^\dagger(3,2)$
from the doublet-triplet $\Phi(3,2)$.

\begin{equation}
i {h^C\over 2} {\bar\Psi}_L(3,2) \times \Psi_R^C(3,1) \cdot
{\tilde \Phi}^\dagger(3,2) + h.c..
\end{equation}
Here vacuum expectation values of $\Phi^0(3,2)$ give rise to
the imaginary off-diagonal (hermitian) elements in the
$3\times 3$ mass matrix, so removing the equal masses of the
charged leptons.

Here the couplings $h^C$ and $h$ are closely related to
the coupling strength $\kappa$ for the family gauge bosons
\cite{Family}.

We wish to note that the last entity is the main basis of the proposed
family collider. Here the coupling $\eta'_1 \mu^+ e^-$ (see below for
explanation) is defined (for the direct production of the mixed family
Higgs $\eta'_1$).

As said in \cite{Origin}, we suppose that, before the spontaneous symmetry
breaking (SSB), the Standard Model does not contain any parameter that is
pertaining to "mass", but, after the SSB, all particles in the Standard
Model acquire the mass terms as it should - we call it "the origin of mass".

In our Standard Model \cite{Hwang417}, we begin with the Standard-Model
Higgs $\Phi(1,2)$, the purely family Higgs $\Phi(3,1)$, and the mixed
family Higgs $\Phi(3,2)$, with the first label for $SU_f(3)$ and the
second for $SU_L(2)$. We need another triplet $\Phi(3,1)$
since all eight family gauge bosons are massive \cite{Family}.

The three Higgs fields sound too many but we are forced to have
them. In fact, the quartic interactions for complex scalar fields
can easily explain the repulsive nature between them, and
the attractive forces can be built up among the related
three complex fields \cite{definition}. They give rise to
that in \cite{Origin}. The "ignition" turns out to come
from the channel in the purely family Higgs $\Phi(3,1)$.

The potential among the three Higgs is given as follows:

\begin{eqnarray}
V_{Higgs} =& \mu^2_2 \Phi^\dagger(3,1) \Phi(3,1) + \lambda
(\Phi^\dagger(1,2) \Phi(1,2)+ cos\theta_P\Phi^\dagger(3,2)\Phi(3,2))^2\nonumber\\
    &  + \lambda(-4 cos\theta_P)
(\Phi^\dagger(\bar 3,2)\Phi(1,2))(\Phi^\dagger(1,2)\Phi(3,2))
  \nonumber\\
  &\lambda
(\Phi^\dagger(3,1) \Phi(3,1)+ sin\theta_P \Phi^\dagger(3,2)\Phi(3,2))^2
    + \lambda(-4 sin\theta_P)
(\Phi^\dagger(\bar 3,2)\Phi(3,1))(\Phi^\dagger(3,1)\Phi(3,2))
  \nonumber\\
     & + \lambda'_2 \Phi^\dagger(\bar 3,1)\Phi(3,1) \Phi^\dagger(1,2) \Phi(1,2)
  + (terms\,\, in\,\, i\delta's\,\, and\,\, in\,\, decay).
\end{eqnarray}
Here the interaction between $\Phi(3,1)$ and $\Phi(1,2)$, as characterized
by the $\lambda'_2$ term, is rather small or vanishes identically.

Exercises in the U-gauge give us what is going on among physical particles.
In the U-gauge, we have
\begin{equation}
\Phi(1,2)= (0,{1\over \sqrt 2} (v+\eta)),\,\, \Phi^0(3,2) = {1\over \sqrt 2}
(u_1+\eta'_1, u_2+ \eta'_2, u_3+\eta'_3 ),\,\, \Phi(3,1) = {1\over \sqrt 2}
(w+\eta',0,0),
\end{equation}
all in columns. The five components of the complex triplet $\Phi(3,1)$ get
absorbed by the $SU_f(3)$ family gauge bosons and the neutral part of
$\Phi(3,2)$ has three real parts left - together making all eight family
gauge bosons massive.

Let us try to fix the notations further; see Ch. 13,
Ref. \cite{Book}. For $\eta'$ going through SSB, we have the following
terms, neglecting the small ${\lambda'}_2$ term,
\begin{equation}
{\mu_2^2\over 2}(\eta'+w)^2+({\epsilon_2\over 4} u_i u_i +{\eta_2\over 4}
u_1^2) (\eta'+w)^2 + {\lambda_2\over 4}(\eta'+w)^4. \nonumber\\
\end{equation}
SSB means that all the linear terms add up to zero, resulting the
change in sign of the mass term, ${1\over 2} (2\lambda_2 w^2 (\eta')^2)$.
(Here, for notations, $\epsilon_2$, $\eta_2$, etc., see \cite{Origin}.)
Note that, for real fields, a factor of ${1\over 2}$ should be factored
out.

The same applies to other SSB fields, even though the original
minus signs are generated by other fields. For example, this
applies for the SM Higgs $\eta$.

In other word, the three "related" scalar (Higgs) fields
$\Phi(1,2)$, $\Phi(3,2)$, and $\Phi(3,1)$ should be "equivalent"
among themselves. The spontaneous symmetry breaking (SSB) is
happening for all of them, actively or passively. It is easy
to see that only one SSB-driving term is enough for all the
three Higgs fields. As for the SSB-driving term (ignition),
we use \cite{Origin} the purely family term, $\mu^2_2
\Phi^\dagger(3,1)\Phi(3,1)$.

From the expressions of $u_iu_i$ and $v^2$, we obtain
\cite{Origin}:
\begin{equation}
v^2 (3 cos^2\theta_P-1) = sin\theta_P cos\theta_P w^2.
\end{equation}
And the SSB-driven $\eta'$ yields
\begin{equation}
w^2 (1-2 sin^2\theta_P) = - {\mu_2^2\over \lambda} +
(sin2\theta_P - tan\theta_P) v^2.
\end{equation}
These two equations show that it is necessary to have the driving
term, since $\mu^2_2=0$ implies that everything is zero. Also,
$\theta=45^\circ$ is the (lower) limit.

What follows below about the masses of $\eta$, $\eta'$, and $\eta'_{1,2,3}$
from \cite{Origin}:

The mass squared of the SM Higgs $\eta$ is $2\lambda cos\theta_P u_i u_i$,
as known to be $(125\,\, GeV)^2$. The famous $v^2$ is the number
divided by $2\lambda$, or $(125\,\,GeV)^2/(2\lambda)$. Using PDG's for
$e$, $sin^2\theta_W$, and the $W$-mass \cite{PDG}, we find
$v^2=255\,\, GeV$. So, $\lambda={1\over 8}$, a simple model indeed.

The mass squared of $\eta'$ is $-2(\mu_2^2-sin\theta_P u_1^2 +
sin\theta_P (u_2^2+u_3^2))$. The  other condensates are $u_1^2= cos\theta_P v^2
+ sin\theta_P w^2$ and $u_{2,3}^2 = cos\theta_P v^2 - sin\theta_P w^2$ while
the mass squared of $\eta'_1$ is $2\lambda u_1^2$, those of $\eta'_{2,3}$ be
$2\lambda u_{2,3}^2$. The mixings among $\eta'_i$ themselves are neglected
in this paper.

There is no SSB for the charged Higgs $\Phi^+(3,2)$. The mass
squared of $\phi_1$ is $\lambda(cos\theta_P v^2 - sin\theta_P w^2) + {\lambda\over 2}
u_i u_i$ while $\phi_{2,3}$ be $\lambda(cos\theta_P v^2 + sin\theta_P w^2)
+ {\lambda\over 2} u_i u_i$. (Note that a factor of ${1\over 2}$ appears
in the kinetic and mass terms when we simplify from the complex case to
that of the real field; see Ch. 13 of \cite{Book}.)

A further look of these equations tells that $3cos^2\theta_P - 1 > 0$ and
$2sin^2\theta_P -1 > 0$. A narrow range of $\theta_P$ is allowed (greater
than $45^\circ$ while less than $57.4^\circ$, which is determined by
the group structure). For illustration, let us choose
$cos \theta_0 = 0.6$ and work out the numbers as follows:
(Note that $\lambda={1\over 8}$ is used.)
\begin{eqnarray}
& 6 w^2 = v^2, \quad -\mu^2_2/\lambda = 0.32 v^2;\nonumber\\
\eta: & m(\eta) =125\, GeV, \quad v^2 = (250\,GeV)^2;
\nonumber\\
\eta': & m(\eta') = 51.03\,GeV, \quad w^2=v^2/6; \nonumber\\
\eta'_1: & m(\eta'_1)= 107\,GeV, \quad u_1^2=0.7333 v^2; \nonumber\\
\eta'_{2,3}: & m(\eta'_{2,3}) = 85.4\,GeV, \quad u_{2,3} = 0.4667 v^2; \nonumber\\
\phi_1:& mass = 100.8\, GeV; \qquad \phi_{2,3}: mass = 110.6\,GeV.
\end{eqnarray}
All numbers appear to be reasonable. Since the new objects need to be
accessed in the lepton world, it would be a challenge for our experimental
colleagues. Note that they are smooth functions of $\theta$ as long as
it is between $45^\circ$ and $57.4^\circ$.

As for the range of validity, ${1\over 3} \le cos^2\theta_P \le {1\over 2}$.
The first limit refers to $w^2=0$ while the second for $\mu_2^2 = 0$.

We may fix up the various couplings, using our common senses. The
cross-dot products would be similar to $\kappa$, the basic coupling of
the family gauge bosons. The electroweak coupling $g$ is
$0.6300$ while the strong QCD coupling $g_s=3.545$; my first guess
for $\kappa$ would be about $0.1$. The masses of the family gauge
bosons would be estimated by using ${1\over 2}\kappa \cdot w$, so
slightly less than $10\,GeV$. (In the numerical example with $cos
\theta_P=0.6$, we have $6 w^2= v^2$ or $w=102\,\,GeV$. This gives
$m=5\,\,GeV$ as the estimate.) So, the range of the family forces,
existing in the lepton world, would be $0.04\,\, fermi$.

For the quark world, or the lepton world, which the $SU_c(3)
\times SU_L(2) \times U(1) \times SU_f(3)$ Minkowski space-time
supports \cite{definition}, the story is also fixed if the so-called
"gauge-invariant derivative", i.e. $D_\mu$ in the kinetic-energy
term $-\bar \Psi \gamma_\mu D_\mu \Psi$, is given for a given
basic unit \cite{Book}.

Thus, we have, for the up-type right-handed quarks $u_R$, $c_R$,
and $t_R$,
\begin{equation}
D_\mu = \partial_\mu - i g_c {\lambda^a\over 2} G_\mu^a -
i {2\over 3} g'B_\mu,
\end{equation}
and, for the rotated down-type right-handed quarks $d'_R$, $s'_R$,
and $b'_R$,
\begin{equation}
D_\mu = \partial_\mu - i g_c {\lambda^a\over 2} G_\mu^a -
i (-{1\over 3}) g' B_\mu.
\end{equation}

On the other hand, we have, for the $SU_L(2)$ quark doublets,
\begin{equation}
D_\mu = \partial_\mu - i g_c {\lambda^a\over 2} G_\mu^a - i g
{\vec \tau\over 2}\cdot \vec A_\mu - i {1\over 6} g'B_\mu.
\end{equation}

For the lepton side, we introduce the family triplet,
$(\nu_\tau^R,\,\nu_\mu^R,\,,\nu_e^R)$ (column), under $SU_f(3)$.
Since the minimal Standard Model does not see the right-handed
neutrinos, it would be a natural way to make an extension of the
minimal Standard Model. Or, we have, for $(\nu_\tau^R,\,
\nu_\mu^R,\,\nu_e^R)$,
\begin{equation}
D_\mu = \partial_\mu - i \kappa {\bar\lambda^a\over 2} F_\mu^a.
\end{equation}
and, for the left-handed $SU_f(3)$-triplet and $SU_L(2)$-doublet
$((\nu_\tau^L,\,\tau^L),\, (\nu_\mu^L,\,\mu^L),\, (\nu_e^L,\,e^L))$
(all columns),
\begin{equation}
D_\mu = \partial_\mu - i \kappa {\bar\lambda^a\over 2} F_\mu^a - i g
{\vec \tau\over 2} \cdot \vec A_\mu + i {1\over 2} g' B_\mu.
\end{equation}
The right-handed charged leptons form the triplet $\Psi_R^C(3,1)$ under
$SU_f(3)$, since it were singlets their common factor $\bar\Psi_L(\bar 3,2)
\Psi_R(1,1)\Phi(3,2)$ for the mass terms would involve the cross terms such as
$\mu\to e$.

In other words, the quarks don't see the family, i.e. $SU_f(3)$, but leptons
see fully the family.

These allow us to write down how the quarks and the leptons enter
in the standard manner.

In our example, the masses for family
Higgs $\eta'$ and $\eta'_{1,2,3}$ are $(50-110)\,\, GeV$ - accessible
only through the lepton world. The implication of the
family gauge theory is in fact a multi-GeV or sub-sub-fermi
gauge theory - the leptons are shielded from this $SU_f(3)$ theory
against the QED Landau's ghost.

The masses of quarks are diagonal, or the singlets in the
$SU_f(3)$ space (in the old-fashion way), those of the
three charged leptons are $m_0 + a \lambda_2 + b \lambda_5 +
c \lambda_7$ (before diagonalization, with real $a$, $b$, an $c$)
and the masses of neutrinos are
purely off-diagonal, i.e. $a' \lambda_2 + b' \lambda_5 + c' \lambda_7$.
This result is very interesting and very intriguing.

Neutrinos oscillate among themselves, giving rise to a lepton-flavor-violating
interaction (LFV). As argued in the other context, there are other oscillation
stories, such as the oscillation in the $K^0-{\bar K}^0$ system, but there is
a fundamental "intrinsic" difference here - the $K^0-{\bar K}^0$ system is
composite while neutrinos are "point-like" Dirac particles. We have standard
Feynmann diagrams for the kaon oscillations but similar diagrams do not exist
for point-like neutrino oscillations - our proposal solves the problem, maybe
in a unique way.

Thinking it through, it is true that neutrino masses and neutrino
oscillations may be regarded as one of the most important experimental
facts over the last thirty years \cite{PDG}.

In fact, certain LFV processes such as $\mu \to e + \gamma$ \cite{PDG},
$\mu + A \to A^* + e$, $e^+ + e^- \to \mu^+ + e^-$, etc., are closely
related to the most cited picture of neutrino oscillations \cite{PDG}. In
recent publications \cite{Hwang10}, it was pointed out that the cross-generation
or off-diagonal neutrino-Higgs interaction may serve as the detailed mechanism
of neutrino oscillations, with some vacuum expectation value of the family Higgs,
$\Phi(3,1)$ and $\Phi^0(3,2)$. So, even though we haven't seen, directly,
the family gauge bosons and family Higgs particles, we already see the
manifestations of their vacuum expectation values.

We suspect that neutrino oscillations and lepton-flavor violating reactions
(or decays) would, eventually, help us to decide, in the great details, the
whereabout of the Standard Model.

Besides the three Higgs, the primary prediction of our Standard Model
is the existence of the force of a new kind - i.e., the family force
mediated by the family gauge bosons. As said above, we could use
${1\over 2}\kappa w$ as an estimate of the mass(es) of the family
gauge bosons. My first guess is for some feeble force - $\kappa=0.1$.
Our numerical example corresponds to $w=102\,\, GeV$, so as to
the family gauge boson mass of $5\,\,GeV$.

The family gauge bosons would then be in the vicinity
of $5\,\, GeV$ or nearby, or the range of $0.04\,\,fermi$.
Or, $0.04 \times 10^{-13}$ cm in the
effective range, between leptons (such as two electrons or an
electron-positron pair) is too short to be detected
for the entire atomic physics or the entire chemistry.

The precision experiments such as $g-2$ would eventually detect the
residual family effects, since the existing $g-2$ calculations
\cite{Kinoshita} is so far the QED calculation and should be
completed by inclusion of other effects with the emphasis on
family gauge bosons. We are looking forward to prospects in this
directions.

Of course, we need to examine the precision part of atomic physics
when the story becomes clear; even though the effects are tiny,
the evolutions usually come from the tiny effects to begin with.

\medskip

\section{The $\mu^+e^-$ Collider}

The idea to have the $\mu^+e^-$ collider at $(80-120)\,GeV$
CM energies is to produce the family Higgs $\eta'_{1,2,3}$
directly. The $\epsilon_{abc}$ coupling in the lepton world
means that so far the $\mu^+e^-$ collider would be only the
available collider.

We have plenty of experience in constructing high-quality
electron beams but may be lack of obtaining a $\mu^+$ beam
in view of the short muon lifetime. There were the proposals
of the $\mu^=\mu^-$ collider. So, the proposal of the
$\mu^+e^-$ collider might not be outrageous.

The cost for the $\mu^+e^-$ collider at, e.g., $120\,GeV$
should be in the ball part of LEP of LEPII previously at
CERN.

The (old) standard wisdom is that, besides the QED
background, you would see nothing in the absence of
the $SU_f(3)$; but, in view of $SU_f(3)$, you should
see $\eta'_{1,2,3}$ and many other family things. Do
you want to gamble on this? I think that, until you
check on this, we don't know where the three
generations (or the family) come from.

If the (new) Standard Model is substantiated, there
are so many things both for theorists and for
experimentalists.

\medskip

\section{Why is the hundred GeV $\mu^+e^-$ collider the only family collider?}

The mixed family Higgs $\eta'_1$ can only be produced directly
by the $\mu^+ e^- \eta'_1$ coupling (in Eq. (2)), thus by
the proposed family collider. The hadron collider, such as
LHC at Geneva, has nothing with it; even the $e^+e^-$
collider could not do it directly. This is why the phenomena
of the three generations are clearly there but the objects
such as $\eta'_1$ have been rather elusive, experimentally.

Of course, if we are satisfied with some indirect evidences,
we could do it in many other ways. For example, we eventually
should see the effects of the family gauge bosons, in messy
environments (since these family bosons couple to the
$e^+e^-$ pair), or through a clear deviation of $g-2$ from the
huge QED-Weak "background". Here so far and what follows, we
assume that the ordering of the couplings, $g_W/g_c \sim
\kappa /g_W$, is approximately valid. To disentangle the
minute effects in the atomic physics environments would
eventually becomes our goal - it is as simple as the QED-Weak
situation.

One would argue against constructing a one-purpose collider,
but the same argument held also for the $e^+e^-$ collider or
for LHC. After all, it would be a nontrivial task to build
the $\mu$ beam that could be used in the game, since the
muon lifetime (at rest) is so short.

\medskip

\section{Are they too narrow to escape the detection?}

Our experience in the detection of weak bosons $W^\pm$ and
$Z^0$ with $GeV$ widths is to use the scanning technique.
Remembering that we assume the coupling strength $\kappa$
to be about one order smaller than the weak coupling
$g_W$, we thus have to be able to fight to see the
ten's $MeV$ in the one-hundred $GeV$ environments - a
resolution of $10^{-5}$ (in energy) is needed
if the same scanning technique could be used.

This also implies that we may have missed important
things if the coupling turns out to be too small.
In that case, we have to invent new detection
techniques in order to unravel more in the Nature.

Of course, there are plenty of techniques available
to us; some of them with miracle invention may work;
etc. The field of knowledge, in theory and in
experiment, seems to be unlimited.

\end{document}